\documentclass[%
superscriptaddress,
bibnotes,
 amsmath,amssymb,
 aps,
 prl, 
longbibliography,
eprint,
floatfix,
reprint
]{revtex4-2}

\usepackage{xcolor}
\definecolor{darkblue}{RGB}{0,0,150}
\definecolor{nightblue}{RGB}{0,0,100}

\usepackage{mathrsfs,dsfont,mathtools,braket,MnSymbol}

\newcommand{\refsub}[2]{\hyperref[#1]{\ref*{#1}#2}}

\usepackage{bm}% bold math
\usepackage[
colorlinks,
citecolor=darkblue,
linkcolor=darkblue,
urlcolor=nightblue]{hyperref}% add hypertext capabilities
\usepackage[english]{babel}
\usepackage[babel,kerning=true,spacing=true]{microtype}

\definecolor{DarkRed}{RGB}{100,0,0}
\definecolor{LightGreen}{RGB}{000,50,0}
\def\({\left(}
\def\){\right)}
\def\prl{\partial}

\newcommand{\beq}{\begin{equation}}
\newcommand{\eeq}{\end{equation}}
\newcommand{\bal}{\begin{aligned}}
\newcommand{\eal}{\end{aligned}}

\begin{document}
\title{
Quantum Critical Dynamics Induced by Topological Zero Modes
}

\author{Ilia Komissarov}
\email{i.komissarov@columbia.edu}
\affiliation{Department of Physics, Columbia University, New York, NY 10027, USA}
\author{Tobias Holder}
\email{tobiasholder@tauex.tau.ac.il}
\affiliation{School of Physics and Astronomy, Tel Aviv University, Tel Aviv, Israel}
%\affiliation{Department of Condensed Matter Physics, Weizmann Institute of Science, Rehovot, Israel}
\author{Raquel Queiroz}
\email{raquel.queiroz@columbia.edu}
\affiliation{Department of Physics, Columbia University, New York, NY 10027, USA}
\affiliation{Center for Computational Quantum Physics, Flatiron Institute, New York, New York 10010, USA}

\date{\today}

\begin{abstract}
We investigate low-frequency ac transport in the Su-Schrieffer-Heeger (SSH) chain with chiral disorder near the topological delocalization transition. Our key finding is that the formation of hybridized pairs of topological domain wall zero modes leads to the anomalous logarithmic scaling of the ac conductivity $\sigma(\omega) \sim \log \omega$ at criticality, and $\sigma(\omega) \sim \omega^{2|\delta|}\log^2 \omega$ away from it.
 Using the combination of real-space renormalization group analysis and qualitative hybridization arguments, we demonstrate that the form of the scaling of ac conductivity at criticality stems directly from the stretched-exponential ($\psi(x)\sim e^{-s\sqrt{x}}$) spatial decay of zero-mode wavefunctions at the critical point.
\end{abstract}

\maketitle

\paragraph{Introduction.---} 

Transport phenomena in disordered topological insulators differ strikingly from those in trivial insulators. For example, some subset of electronic states may become delocalized even in low-dimensional systems --- a prominent example being the quantum Hall plateau transition \cite{evers_anderson_2008, qh1, qh2, qh3}, which, despite many years of research, remains poorly understood \cite{sarma}. A simpler and more tractable model that displays similar behavior is the Su-Schrieffer-Heeger (SSH) chain \cite{ssh} with chiral disorder, i.e., with disorder in hopping elements only. In this model, exactly one mode at zero energy becomes delocalized when the geometric averages of even and odd hopping elements are equal \cite{theodorou_extended_1976, altland_topology_2015}.

Many aspects of quantum critical behavior in this model and the ones in the same universality class were extensively studied \cite{tr1, tr2, soukoulis_off-diagonal_1981, zhang_anatomy_2023, balents_delocalization_1997, gogolin_electron_1988}, and observed experimentally in disordered atomic wires \cite{meier}. The renormalization-group flow equations were solved analytically \cite{fisher1, fisher2}, providing access to exact results, such as the low-energy density of states and critical exponents. For the continuous $1D$ fermion model with random mass, characterized by the same fixed point, it has been shown that at criticality, charge carriers exhibit a logarithmically slow spreading of probability density \cite{bouchaud_classical_1990, bagrets_sinai_2016}. 

\begin{figure}[t!]
    \centering
    \includegraphics[width=0.45\textwidth]{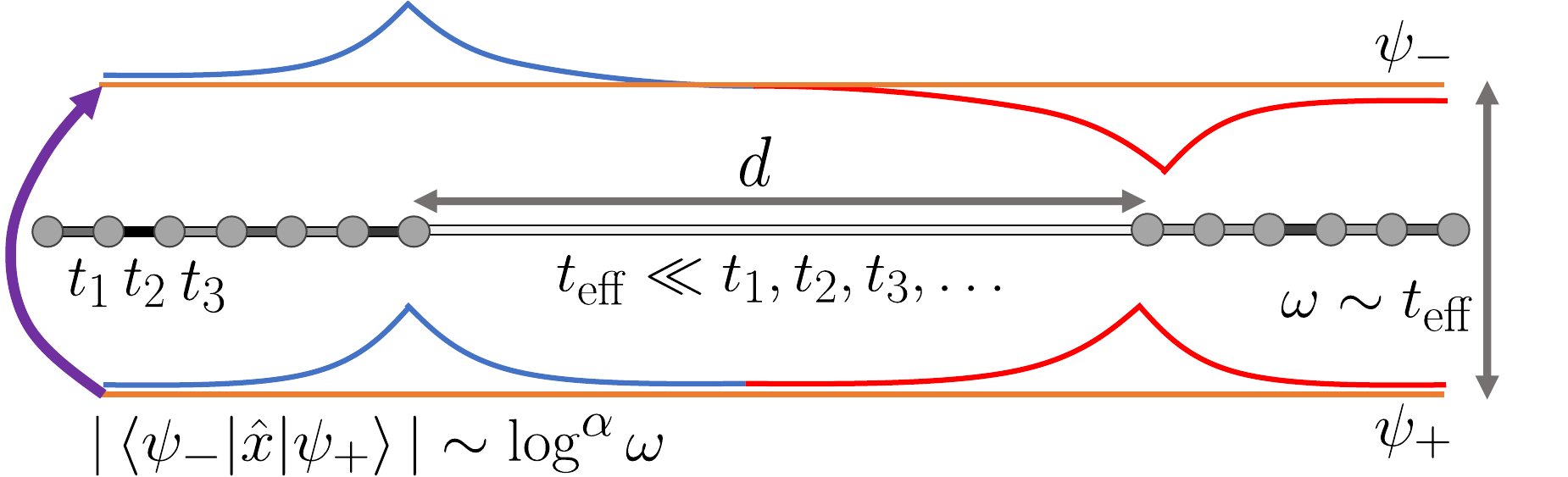}
    \caption{Hopping disorder in the SSH chain creates isolated rare regions characterized by extremely small effective tunneling probability $t_{\rm eff}$. Two zero modes appearing on both sides of the link $t_{\rm eff}$ hybridize away from zero energy with energy splitting $\omega \sim t_{\rm eff} \sim e^{-\sqrt[\alpha]{d/\xi}}$, where $\xi$ encodes the typical size of individual zero modes and both $\alpha=1$ and $\alpha=2$ are possible in this model. The transition amplitude between these states in a time-dependent electric field is $|\braket{\psi_-| \hat x| \psi_+}| \sim d \sim \log^\alpha \omega$, implying an insulator for $\alpha = 1$ with ac conductivity following a modified Mott-Berezinskiy law \eqref{soffcrit}, and, for $\alpha=2$, a glassy metal characterized by activated dynamical scaling \eqref{scrit}.} 
    \label{plot1}
\end{figure}

In this work, we clarify the nature of the low-energy states that dominate the ac conductivity in this model. Our analytical approach follows the framework introduced by Mott \cite{mott_electronic_2012}, who proposed that special rare resonating states dominate the ac conductivity in the Anderson model. We demonstrate that, in the SSH chain with chiral disorder, we can identify a relevant subset of states for which the matrix element of the dipole operator diverges at small $\omega$. We show that the role of such states is played by bonding ($\psi_+$) and anti-bonding ($\psi_-$) orbitals formed by hybridizing pairs of topological zero modes (see \autoref{plot1}). The dipole transition amplitude between $\ket{\psi_+}$ and $\ket{\psi_-}$ at low $\omega$ behaves as $\log \omega$ away from criticality and $\log^2 \omega$ exactly at the critical point due to the exponential and the stretched-exponential decay of the wavefunctions of zero modes \cite{balents_delocalization_1997}.

Our approach leads to the logarithmic scaling of low-frequency ac conductivity at criticality
\begin{equation}
    \sigma_{\rm crit}(\omega) \sim \log (\Omega/\omega) \, ,
    \label{scrit}
\end{equation}
where $\Omega$ is a high-energy reference scale. This form of ac conductivity is highly atypical: as the frequency $\omega$ is decreased, we would expect the number of states participating in transport to decrease polynomially due to Pauli blocking. This would lead to insulating behavior as it occurs in the chain with purely on-site disorder (Anderson model), where $
\sigma_{\rm MB}(\omega) \sim \omega^2 \log^2 \omega \, ,$
according to the Mott-Berezinskiy law \cite{mott_electronic_2012, abrikosov_conductivity_1978, khalatnikov_kinetics_1996}. Instead, in the case of the SSH chain with chiral disorder and at criticality, the divergence of the density of the in-gap states induced by the fluctuations in the winding number introduced by impurities (Dyson singularity \cite{evers_anderson_2008, eggarter}) compensates for the Pauli blocking effects, and the dc conductivity diverges. This divergence is, however, only logarithmic in frequency, which indicates the tunneling regime of electronic propagation with the vanishing Drude weight, in contrast to the ``free acceleration'' $\sigma(\omega) \sim \omega^{-1}$ form \cite{drude} typical in conventional metals. Therefore, the appearance of \eqref{scrit} highlights a remarkable situation in which the topology almost exactly counteracts the localizing effects of disorder, giving rise to logarithmically slow ``glassy'' behavior of the ac conductivity. We note that using chiral disorder is crucial for this behavior, as on-site impurities would lower the symmetry to the topologically trivial Anderson model without a divergent density of states.

Using  similar hybridization arguments, we also argue that away from the critical point, the form of the ac conductivity in the SSH chain with chiral disorder assumes the form
\begin{equation}
    \sigma(\omega) \sim \omega^{2 |\delta|} \log^2 \omega \, ,
    \label{soffcrit} 
\end{equation}
with $\delta$ parameterizing the distance to the critical point 
~\footnote{See supplemental material, where we explain the main ingredients of the real-space renormalization group in one dimension, and give instructions on how to apply the scaling analysis to other models.}. Thus, in the $\delta \neq 0$ case, the dc conductivity vanishes since the divergence in the density of zero modes is milder than in the critical case and insufficient to compensate for the reduction of low energy excitations with energy $\omega$. Both Eq. \eqref{scrit} and Eq. \eqref{soffcrit} are in agreement with previous studies on spin chains \cite{motr}, provided that the correspondence $2 |\delta| = z^{-1}$ is taken into account \cite{McKenzie}, where $z$ is the dynamical critical exponent, which diverges at the transition. Dynamical properties similar to Eq.~\eqref{scrit} and Eq.~\eqref{soffcrit} were also obtained previously near superconductor-metal transitions \cite{scmet}, and many-body localized phases \cite{Gopalakrishnan_2015}. 

Our work complements these earlier results by relating the scaling of ac conductivity in Griffiths systems to hopping conduction phenomena in disordered insulators studied by Mott and Berezinskii \cite{mott_electronic_2012, khalatnikov_kinetics_1996}. To establish this relation, a precise understanding of the hybridized topological zero modes is central, from which we obtain several new properties regarding the scaling of the ac conductivity and the role of topology in transport. Namely, we connect the decay of the critical zero-mode states (i.~e., the scaling of the matrix element of the position operator) with the energy separation between the states (cf. also \cite{annica}). The result is a low-disorder scaling of ac conductivity $\sigma_{\rm crit}(\omega) \sim s^{-2} \log (s^2/\omega)$ expressed in terms of the parameter $s$ characterizing the disorder strength. Unlike conventional real-space renormalization-group methods, the more general Mott framework, which relies solely on the sufficient spatial separation of hybridized pairs \cite{kirsch_mott_2003}, enables us to study transport properties at relatively high energy scales analytically. In particular, we demonstrate a smooth crossover between the regimes Eq.~\eqref{scrit} and Eq.~\eqref{soffcrit}. This implies that observing the activated dynamical scaling in Eq. \eqref{scrit} does not require the fine-tuning of parameters to criticality, and can be performed in an off-critical system at sufficiently high frequencies $\omega$. Lastly, we find that ac conductivity receives a topological enhancement~\cite{windingr}.

In the following, we study the low-frequency dynamics of the SSH chain with bond dilution as an intuitive example that elucidates the role of zero-mode resonances in transport. The case of continuous bond disorder distributions follows from similar considerations. Finally, we present analytical and numerical evidence that, at criticality, the transition amplitude between the Mott resonances scales as $\log^2 \omega$, leading to the activated scaling of the low-frequency ac conductivity, as given in Eq.~\eqref{scrit}. 

\paragraph{AC conductivity in the bond-diluted SSH chain.---}
\label{sec1}
To gain intuition about the role of zero modes in the low-frequency transport of disordered topological insulators, we first consider an example where bonds are removed randomly from an SSH chain \cite{ssh}
\begin{equation}
    \hat H_{\rm SSH} = \sum_i t_i' \ket{i,\rm B} \bra{i,\rm A}  + t_i \ket{i+1,\rm B} \bra{i,\rm A} + {\rm h.c.}\, ,
    \label{sshhd}
\end{equation}
where $t_i$ and $t_i'$ take either their unperturbed values $t$ and $t'$ with probability $1-p$ or 0 with probability $p$. Since our goal is accessing the low-frequency transport at zero energy, we are interested in the structure of states around $E=0$. We assume $t'<t$.

When $p > 0$, the chain breaks down into isolated segments among which only the ones terminating with $t'$ on both sides contribute to the low-frequency ac conductivity. These regions host pairs of zero modes in between which optical transitions are allowed in the sub-gap regime $\omega \ll |t-t'|$. To see this, consider a topological segment of length $d$. If $d$ is large enough, the zero modes at its ends hybridize with each other without mixing with other states due to the finite energy gap. From perturbation theory (see \cite{Note1}), the energy associated with this hybridization is proportional to the exponentially small overlap between the zero modes, $\omega = \Omega  e^{-d/ \xi}$, with $\xi$ encoding the typical size of the zero mode which is proportional to its localization length $\xi_0 = \log^{-1}|t/t'|$, see \cite{asboth}. It follows that the long topological segment hosts low-lying in energy resonant bonding and anti-bonding states $\ket{\psi_{\pm}} = (\ket{\psi_{\rm L}} \pm \ket{\psi_{\rm R}})/\sqrt{2}$ of the zero modes at the left $\ket{\psi_{\rm L}}$ and right $\ket{\psi_{\rm R}}$ ends of the section. The dipole matrix element between the resonances takes the form
\begin{equation}
\begin{aligned}
    |\braket{\psi_+|\hat x|\psi_-}| \sim |\braket{\psi_{\rm L}|\hat x|\psi_{\rm L}} - \braket{\psi_{\rm R}|\hat x|\psi_{\rm R}}|  \sim & \, d \sim \xi_0 \log \omega \, ,
    \label{x}
\end{aligned}
\end{equation}
while the density of such states per unit length arises from all topological sections of arbitrary length $d$, appearing with probability $(1-p)^d$. That is,
\begin{equation}
    \nu_{\rm topo}(E) \sim \sum_d (1-p)^{d} \delta\left(E - \Omega e^{-d/\xi} \right) \sim |E|^{-1 +2 |\delta|}\, , 
    \label{dos}
\end{equation}
where in this model $|\delta|= (\xi/2) |\log(1-p)|$  {diverges at criticality}. Similar algebraic behavior of the density of states is well-known in the theory of disordered magnets \cite{vojta_phases_2013}, and a direct mapping to the $1D$ random-bond Ising model can be formally established: see supplementary information (SI) in Ref.~\cite{Note1}. Continuing this analogy, the model given by Eq.~\eqref{sshhd} with $0<|\delta| < 1/2$ is in the so-called Griffiths phase with the hybridized pairs of zero modes playing the role of the pairs of domain walls which form the rare magnetized regions \cite{mccoy, vojta_phases_2013, grif}.

\begin{figure*}[t!]
    \centering
    \includegraphics[width=1\textwidth]{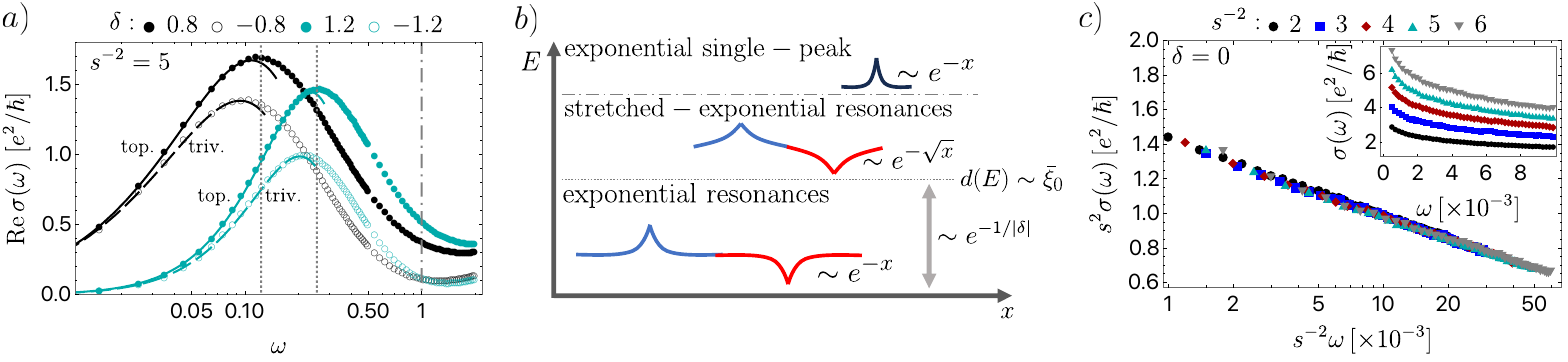}
    \caption{$a)\,$Dynamical conductivity in the SSH chain with chiral box disorder at half filling for different values of the parameter $\delta$, which encodes the distance to the topological phase transition. The numerical data is fitted to $\sigma(\omega) = c_1 \omega^{2 |\delta_{\rm fit}|} \log^2 (c_2/\omega)$, and $|\delta_{\rm fit}|= 0.78$, $0.78$, $1.23$, $1.24$ indicate a good agreement with the actual values of dimerization $\delta$, confirming our expression for ac conductivity in this regime \eqref{soffcrit}. The crossover to the critical $\sigma(\omega) \sim \log \omega$ regime occurs beyond a certain $\delta$-dependent frequency indicated by the dotted lines. Note that ac conductivity is enhanced for $\delta>0$, corresponding to the topological phase in the model. $b)$ Schematic representation of the crossover between different types of states found in the SSH model with bond disorder. When $\delta \neq 0$, the closest to $E = 0$ states are the hybridized exponential modes. Immediately above it lie the stretched-exponential double-peak wavefunctions arising from the surviving critical regions. At even higher energies, regular exponentially localized states are found. $c)$ Same as $a)$ for the critical value $\delta = 0$ and different disorder strengths $s$. The curves exhibit good data collapse after rescaling both the frequency and the conductivity by $s^2$. The unscaled curves are shown in the inset. Importantly, the $\log \omega$ law holds at all disorder strengths, consistent with \eqref{sigcrit}. The simulations were performed for chains of $10^3$ sites and $10^6$ disorder configurations.}
    \label{plot2}
\end{figure*}

Using Eqs.~\eqref{x} and \eqref{dos}, we estimate the ac conductivity at low frequencies at half filling through the Kubo-Greenwood formula \cite{mott_electronic_2012} 
\begin{equation}
    \sigma(\omega) =  \frac{\pi e^2 \omega}{\hbar} \sum_{E_n \leq E_{\rm F} \leq E_m} |\braket{n|\hat x|m}|^2 \delta (E_m - E_n - \omega)  \, .
    \label{kgformula}
\end{equation}
For $\omega \ll |t-t'|$, Eq.~\eqref{kgformula} reduces to the sum over all transitions between the $\psi_{\pm}$ pairs of resonances created by hybridizing zero modes
\begin{equation}
\begin{aligned}
    \sigma(\omega) & \sim \omega \int_0^\omega d E\, \nu(E) |\braket{\psi_{E}|\hat x|\psi_{-E}}|^2 \delta(2 E - \omega) \\ & \sim \omega^{2 |\delta|} \log^2 \omega \, ,
\end{aligned}
    \label{acdelta}
\end{equation}
as stated in Eq.~\eqref{soffcrit}.

\paragraph{AC conductivity away from criticality.---}
\label{sec12}
We now argue that conductivity retains the form of Eq.~\eqref{soffcrit} for a continuous distribution of disordered bonds. To this end, consider the model defined in Eq.~\eqref{sshhd}
with $t_i=t-\epsilon_i$, $t_i' = t' - \epsilon_i$, where $\epsilon_i$, $\epsilon_i'$ are drawn from any smooth distribution. For the numerical simulations, we assume it is uniform over the interval $[-W/2, W/2]$. 

We use real space renormalization group (RSRG) \cite{fisher1, fisher2}, to obtain the properties of low-energy states in this case~\cite{Note1}. This method consists of identifying the largest hopping parameters and removing them from the chain using second-order perturbation theory, thus lowering the energy cutoff $\Omega$, down from its initial value of $\Omega_0 = {\rm max}[t_i,t_i']$ (without loss of generality, we fix $\prod_i t_i'< \prod_i t_i$). 
As this procedure is continuously repeated, the remaining effective hoppings $t'_{\mathrm{eff}}$ become broadly distributed $P(t'_{\mathrm{eff}}) \sim \exp [-2 |\delta|  \frac{\log(\Omega/t'_{\mathrm{eff}})}{(\Omega_0/\Omega)^{2|\delta|}} ]$, 
%$P(t'_{\mathrm{eff}}) \sim (\Omega/t'_{\mathrm{eff}})^[- 2 \delta \exp\{-2\delta \log(\Omega_0/\Omega)\} ]$
where the staggering parameter $\delta$ is related to the typical localization length at zero energy $\xi_0$ as $\delta=\llangle \kappa \rrangle s^{-2}$, where  
\begin{equation}
    \xi_0^{-1} = |\llangle \kappa \rrangle|, \,~ s^2=a(\llangle \kappa^2 \rrangle - \llangle \kappa \rrangle^2), \,~ 
    \kappa \equiv a^{-1}\log \left|\frac{t - \varepsilon}{t' - \varepsilon'}  \right| \, .
    %\kappa \equiv a^{-1}\log \left|\frac{t' - \varepsilon'}{t - \varepsilon}  \right| \, .
    \label{kappas} 
\end{equation}
Here, $\llangle{\cdot}\rrangle$ denotes the disorder average, and $a$ is the lattice constant. We assign the hoppings $t_i$ length $a$ ($L(t_i) = a$) and bonds $t_i'$ zero length ($L(t_i') = 0$), such that $\delta > 0$ corresponds to the topological and $\delta < 0$ to the trivial phases, distinguished by the values of the disorder-averaged winding number \cite{windingr}.  {Note that the criticality in this model corresponds to $\delta = 0$.}

The broad character of the distribution $P(t'_{\mathrm{eff}})$ allows us to apply the Mott hybridization argument to the pairs of states connected by the effective $t'_{\mathrm{eff}}$ hoppings (cf. \autoref{plot1}), since the bonds $t_{\rm eff}'$ \textit{close in real space} are statistically likely to host pairs of zero modes that are very \textit{far apart in energy}. Recalling that the zero modes are exponentially localized and repeating the arguments leading to Eq.~\eqref{x}, we conclude that Eq.~\eqref{x} is also valid for the continuous distribution of bonds. A more rigorous way to obtain Eq.~\eqref{x} using the distribution $P(t'_{\mathrm{eff}})$ is shown in \cite{Note1}. Moreover, considering that the low-energy density of states retains the form of Eq.~\eqref{dos}, $\nu(E) \sim |E|^{2 |\delta|-1}$ \cite{evers_anderson_2008}, we conclude that Eq.~\eqref{soffcrit} for the conductivity remains valid for the case of the continuous distribution of bonds. We also add that the contribution from the optical transitions, which are asymmetric in energy --- absent in the previously discussed case of bond-dilution [Eq.~\eqref{sshhd}] --- does not affect the result as it decays faster than $\omega^{4 |\delta|}$ at low frequencies and therefore subdominant \footnote{
Since the matrix element of the position operator for asymmetric in energy transitions is regular in $\omega$, we estimate from \eqref{kgformula} that $\sigma_{\rm asym}(\omega) \lesssim \omega \int_0^{\omega} dE \, \nu(E)	\nu(E+\hbar \omega) \sim \omega^{4 |\delta|}$}. 

We verify the validity of Eq.~\eqref{soffcrit} by computing $\sigma(\omega)$ numerically using the Kubo-Greenwood formula Eq.~\eqref{kgformula} for the model of Eq.~\eqref{sshhd}  with box disorder of $\epsilon_i, \epsilon_i'$ and different values of the staggering parameter $\delta$. As shown in ~\autoref{plot2}a, Eq.~\eqref{soffcrit} provides reasonable estimates for the numerical data at small frequencies. We also observe that the ac conductivity is enhanced on the topological side due to the increased characteristic length of the low-energy impurity resonances at $\delta > 0$, contributing to $\sigma(\omega)$ via the matrix element of the position operator (see Eq.~\eqref{kgformula}). While one would expect such topological enhancement to affect only transport originating at the lattice scale, we find that it persists for hopping transport over much longer length scales.

\paragraph{AC conductivity at criticality.---}
\label{sec2}
\begin{figure}[t!]
        \centering
        \includegraphics[width=0.4\textwidth]{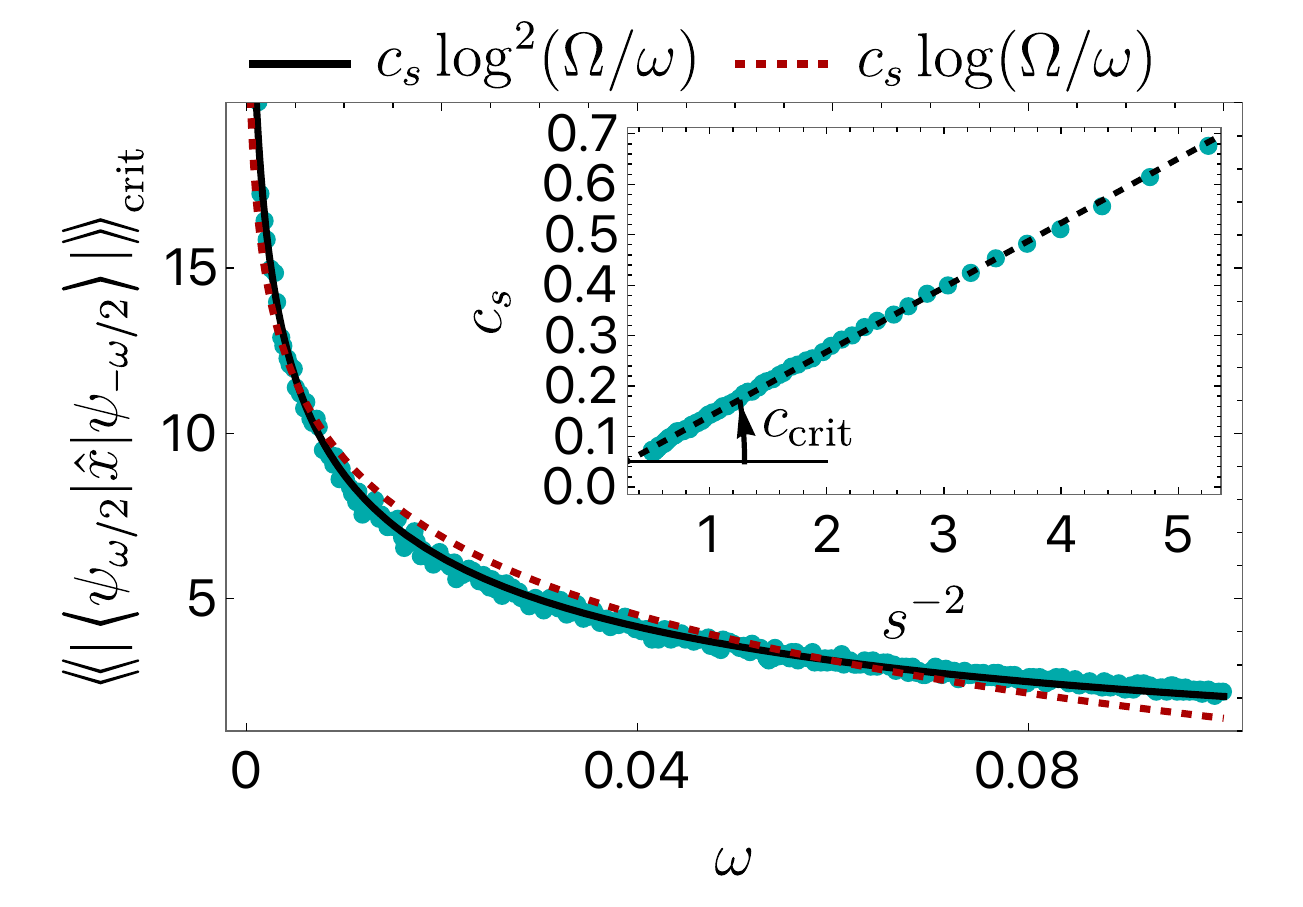}
    \caption{Dipole transition amplitude between states separated by $\omega$ with disorder strength $s^2  = 0.3$, fit to $\log^2 \omega$ (black line) and $\log \omega$ (red dashed line). The best data fit to $\log^2 \omega$ is evidence for stretched exponential behavior of the zero-energy wavefunction at criticality, justifying the result \eqref{logsq}. Inset: coefficient $c_s$ as a function of disorder strength $s$, extracted from fitting $\log^2 \omega$. The predicted relation $c_s = c_{\rm crit}s^{-2}$  in  \eqref{logsq} is established with $c_{\rm crit} \simeq 0.13$. The calculation used $10^3$ sites at criticality ($t=t'=1$), chiral box disorder, and averaging over $10^4$ configurations.}
    \label{plot3}
\end{figure}
We now consider the low-frequency behavior of the ac conductivity in the SSH chain with chiral disorder \eqref{sshhd} at criticality, $t=t'$, with $\epsilon_i$ and $\epsilon_i'$ now drawn from the same disorder distribution, such that the condition $\prod_i t_i = \prod_i t_i'$ is satisfied. At half-filling, this model hosts a single delocalized state at $E=0$ ($\xi_0 = \infty$) \cite{theodorou_extended_1976}. 

Similarly to the off-critical case in the previous section, the low-frequency ac conductivity is dominated by the transitions between isolated bonding and anti-bonding Mott resonances. In this case, all renormalized $t_i,t_i'$ hoppings become broadly distributed with the probability $P(t_{\rm eff}) \sim \exp[- \frac{ \log (\Omega/t_{\rm eff})} {\log(\Omega_0/\Omega)}]$ as the energy cutoff is lowered, which suppresses hybridization events between more than two zero modes, thus again justifying the application of the Mott-type pairwise hybridization argument. In the critical case, the low-energy resonances arise on either $t_{\rm eff}$ or $t_{\rm eff}'$ bonds.

The important aspect that distinguishes the critical case from the off-critical is the stretched-exponential decay of the zero mode wavefunctions at large distances $\psi_0(x) \sim e^{-s \sqrt{x}}$: this unusual spatial dependence is due to the exponent of the wavefunction undergoing a random walk, see SI~\cite{Note1}. Following the same argument leading to Eq.~\eqref{x}, we postulate the transition amplitude between the bonding and antibonding critical Mott resonances in the form
\begin{equation}
    \llangle |\braket{\psi_{\omega/2} | \hat x| \psi_{- \omega/2}}| \rrangle_{\rm crit} \equiv c_{\rm crit} s^{-2} \log^2 (\Omega /\omega)  \, .
    \label{logsq}
\end{equation}
This dependence is confirmed by numerical simulations, as shown in \autoref{plot3}, with the fit to $\log^2 \omega$ agreeing exceptionally well with the data.

Combining the result of Eq.~\eqref{logsq} and the form of the low-energy density of states, given by the Dyson singularity $\nu_{\rm crit}(E) = s^2 [|E| \log^3 \left( \Omega /|E| \right)]^{-1}$~\cite{evers_anderson_2008, eggarter}, we obtain the leading low-frequency asymptote of the ac conductivity at $E_{\rm F} = 0$ via the summation over transitions symmetric in energy relative to $E_{\rm F}$, as in Eq.~\eqref{acdelta}:
\begin{equation}
    \sigma_{\rm crit}(\omega)  \sim  s^{-2} \log (\Omega/\omega) \, .
    \label{sigcrit}
\end{equation}
% \begin{equation}
% \begin{aligned}
%     \sigma_{\rm crit}(\omega) & \sim \omega \int_0^\omega dE\, \nu_{\rm crit}(E) |\braket{\psi_{-E}|\hat x|\psi_E}|_{\rm crit}^2 \delta(2E - \omega) \\
%     & \sim s^{-2} \log \omega \, .
%     \label{sigcrit}
% \end{aligned}
% \end{equation}

The scaling above is valid at any disorder strength, while the more specific form $\sigma_{\rm crit} \sim s^{-2} \log (s^2/\omega)$ is established at low disorder (for $s \ll 1$ when $\Omega \sim s^2$~\cite{Note1}). The validity of Eq.~\eqref{sigcrit} and scaling with $s$ at low disorder are confirmed numerically in \autoref{plot2}c.

Interestingly, the logarithmic scaling \eqref{sigcrit} is also found within a certain frequency range away from criticality (cf. \autoref{plot2}a). This behavior is due to the finite critical topological segments, which are short enough to stay unaffected by the finite localization length $\bar \xi_0 \sim |\delta|^{-2}$ \footnote{In this model, the \textit{average} localization length $\bar \xi_0$ is more divergent at criticality than the \textit{typical} localization length $\xi_0$ \cite{evers_anderson_2008}, which scales merely as $|\delta|^{-1}$ near the critical point, as follows from \eqref{kappas}.}, and hence remain in the form of critical stretched-exponential resonances (a detailed representation of different types of states in the SSH chain with chiral disorder is found in ~\autoref{plot2}b). The crossover occurs when the frequency satisfies $\log^2 \omega \sim \bar \xi_0$, or $\omega \sim e^{-1/|\delta|}$. This argument clarifies the apparent mismatch between \eqref{scrit} and \eqref{soffcrit} at $\delta = 0$; when the staggering is decreased, the lower edge of the region hosting stretched-exponential resonances indicated with dotted lines in \autoref{plot2}a,b approaches $E = 0$, and at any finite $\omega$, a crossover between the two regimes occurs. 

It is instructive to point out that Eq.~\eqref{scrit} is consistent with the known result that the dc conductivity in this model grows as a square root of the size of the system $\sigma(L) \sim \sqrt{L}$ \cite{evers_anderson_2008}. In a finite chain, the transport at the lowest possible $\omega$ occurs due to the tunneling between the two largest resonances of the size $L$, while, according to Eq.~\eqref{logsq}, $L \sim \log^2 \omega$.

 {
Lastly, we note that the divergence of the critical dynamical exponent $z = (2 |\delta|)^{-1}$ is also manifest in the behavior of the dc conductivity as a function of temperature in this model~\cite{Note1}
\begin{equation}
\sigma(T) \sim  \exp\left[-(T/T_0)^{\frac{1}{1+z}}\right]\, ,
\end{equation}
where $T_0$ is a constant reference value. This implies that at the criticality, the transport regime changes from thermally activated to polynomial in $T$, providing a sharp signature in the variable-range hopping exponent, routinely measured in experiments \cite{shk}.}

\paragraph{Conclusion.---}
\label{sec3}
Studying the hybridization of topological zero modes in the SSH chain with chiral disorder, we demonstrated the emergence of a sub-diffusive metallic phase characterized by an activated scaling of the ac conductivity [Eq.~\eqref{scrit}]. Remarkably, this behavior is a consequence of the type of wave function decay \textit{only} - since the latter determines both the dipole transition amplitude between the Mott resonances and the scaling of the density of states entering the expression for the conductivity in Eq.~\eqref{acdelta}. The latter statement comes from the realization that the scaling $\psi_0 \sim e^{-s \sqrt{x}}$, according to our hybridization argument, fixes the length of the typical critical state at energy $\omega$ to be $d(\omega) \sim \log^2 \omega$. This expression, in turn, dictates the typical number of modes with $|E| < \omega$, i.e., $n(\omega) \sim d^{-1} (\omega)\sim \log^{-2} \omega$, and therefore, the Dyson singularity is reproduced $\nu(E) \sim d n(E)/dE \sim |E|^{-1} \log^{-3} |E|$.

We showed that the argument originally introduced by Mott to determine the ac conductivity of Anderson insulators can be extended to the critical sub-diffusive metal phase of the SSH chain, which is described by the infinite randomness fixed point. In this case, the dynamics is dominated by the pairwise hybridization of zero modes, which gives rise to the Dyson singularity and enhances transport at low frequencies. By connecting hopping conductivity with the phenomenon of activated dynamical scaling in Griffiths phases \cite{motr}, we open a pathway towards observing the tunneling metallic behavior in Eq. \eqref{soffcrit} in higher-dimensional disordered insulators with suitable divergence in the density of states.

The recognition of the crucial role of topological in-gap states at criticality opens a new perspective on the microscopic nature of disorder-induced topological phase transitions, such as the quantum Hall plateau transition or critical delocalization in disordered Chern insulators. Indeed, in these systems, similarly, only a single state is delocalized at criticality.  {The states away from the critical inherit its multi-fractal properties by continuity in $E$, with ac conductivity $\sigma(\omega)$ arising due to the optical transitions between these states and carrying the information about the critical behavior.} A crucial distinction of quantum Hall criticality is that the typical wavefunction of the delocalized state is algebraically and not stretched-exponentially localized \cite{evers_anderson_2008}. However, rare region arguments have been successfully applied to algebraically decaying impurity wavefunctions, e.g., in Weyl semimetals \cite{pixley_rare_2021}.

Since the appearance of localized in-gap states around strong impurities is a ubiquitous feature of topological phases in any dimension~\cite{queiroz2024ring}, we anticipate that their resonant hopping will also play a crucial role in determining the existence and dynamics at topological criticality in higher spatial dimensions.

\begin{acknowledgments}
\emph{Acknowledgments ---} We thank Gil Refael for insightful discussions. This work is supported by the National Science Foundation under Award No. DMR-2340394 and the Schwinger Foundation. T.H.\ acknowledges financial support by the European Research Council (ERC) under grant QuantumCUSP (Grant Agreement No. 101077020). 
\end{acknowledgments}

% \section*{Data Availability}
% The data that support the findings of this study are available from the corresponding author upon request.

% \section*{Competing interests}
% The Authors declare no competing interests.

\onecolumngrid
\newpage
\appendix
\begin{center}
\textbf{\large Supplementary information for\\ ``Quantum Critical Dynamics Induced by Topological Zero Modes"}
\\[6pt]
Ilia Komissarov, Tobias Holder, Raquel Queiroz
\end{center}

\stepcounter{page}

\newcommand{\hh}{\textcolor{black}}
\newenvironment{hhc}{\par\color{black}}{\par}
\newenvironment{hbb}{\par\color{black}}{\par}

\newcommand{\hhb}{\textcolor{black}}

\newcommand{\hb}{\textcolor{black}}

\newcommand{\bh}{\textcolor{black}}
\renewcommand{\thefigure}{S\arabic{figure}}

\renewcommand{\thetable}{S\arabic{table}}
\renewcommand{\tablename}{Supplementary Table}

\renewcommand{\thesection}{S\arabic{section}}

\renewcommand{\theequation}{S\arabic{equation}}

% \tableofcontents

% \renewcommand{\thefigure}{\arabic{figure}}
% \renewcommand{\figurename}{Supplementary Figure}
% \renewcommand{\tablename}{Supplementary Table}
% \setcounter{figure}{0}

In this Supplementary Information, we first review the properties of the zero-energy solutions in the SSH chain with chiral disorder. We further analytically address the hybridization of topological zero modes in the bond-diluted SSH chain. We then review certain applications of the real-space renormalization group in the SSH chain with chiral disorder. Additionally, we elaborate on the mapping between the SSH chain model and the random-bond Ising model, as well as the random-mass $1D$ fermion model in a continuum. Lastly, we obtain the temperature dependence of the dc conductivity in the bond-disordered SSH chain.

\section{Notations and conventions} 
The lattice constant $a$ is set to unity, unless stated otherwise. We use $E$ and $E_{\rm F}$ to denote the reference value of the energy and the Fermi energy, respectively. The quantity $\omega$ bears a meaning of the energy difference, hybridization energy, or the frequency of the applied ac electric field expressed in units of energy, depending on the context. With $d(E)$ we denote the characteristic spatial extent of states at energy $E$. We use the notation $\llangle\ldots\rrangle$ for the disorder average, reserving $\braket{\ldots}$ for matrix elements of operators. For the numerical simulations performed for the SSH chain with chiral disorder, we fix the sum of the inequivalent hoppings to a constant value $t+t' = 2$: the remaining free parameters of the model are therefore the difference $t-t'$ and the disorder strength $W$. The edge states are removed from the spectrum if present. 

\section{Wavefunctional decay of zero modes in the SSH chain with chiral disorder}
\label{appa}

In this section, we review the properties of the zero-energy states in the SSH chain with chiral disorder, showing that they decay exponentially at long distances when the system is away from criticality, and stretched-exponentially (as $\psi \sim e^{-s \sqrt{x}}$) at the critical point.

We consider the Schrodinger equation $\hat H_{\rm SSH} \ket{\psi} = 0$ with the Hamiltonian
\begin{equation}
    \hat H_{\rm SSH} = \sum_i t_i' \ket{i,\rm B} \bra{i,\rm A}  + t_i \ket{i+1,\rm B} \bra{i,\rm A} + {\rm h.c.} \, .
    \label{sshhda}
\end{equation}
Using the real-space representation
$\ket{\psi} = \sum_i \left( \psi_{i \rm A} \ket{i, \rm A} + \psi_{i \rm B} \ket{i, \rm B} \right)$, we obtain
\begin{equation}
\begin{aligned}
    t_i' \psi_{i\rm A} + t_i \psi_{i-1\rm A} & = 0 \, , \\ 
    t_i' \psi_{i\rm B} + t_i \psi_{i+1\rm B} & = 0 \, .
\end{aligned}
\end{equation}
By iterating the above expressions, the growth(decay) rates of the zero modes localized on A and B sub-lattices are found
\begin{equation}
    \xi_{0\rm B}^{-1} = -\xi_{0\rm A}^{-1} = \frac{1}{N} \log \left| \psi_{N\rm A}/\psi_{1\rm A} \right|  = \frac{1}{N} \sum_i^N \log \left|   t_i'/t_i\right| \, .
    \label{xiavg}
\end{equation}
Since the sum of the random quantities on the right-hand side is self-averaging, we can introduce the typical localization length of the mode localized on the B sub-lattice as defined by the law of large numbers
\begin{equation}
\xi_{0}^{-1} = \left \llangle \log \left|   t'_i /t_i\right| \right \rrangle \, .
\end{equation}
If the expression on the right-hand side is finite, the zero mode decays exponentially on average. If, on the other hand, the right-hand side vanishes, the exponent of the wavefunction, as we can see from \eqref{xiavg}, undergoes a random walk with zero mean. The spatial decay is stretched-exponential, and characterized by a length scale $s$:
\begin{equation}
    \left\llangle |\log \psi_N|^2 \right\rrangle - \left\llangle |\log \psi_N| \right\rrangle^2 = \sqrt{\left\llangle \left( \sum_{i=1}^N \log \left| t_i'/t_i \right| \right)^2 \right\rrangle} = s \sqrt{N} \, , \qquad s = \llangle \log^2|t_i'/t_i| \rrangle - \llangle \log|t_i'/t_i| \rrangle^2 \, .
\end{equation}
Therefore, we recover the average stretched-exponential zero mode behavior $\psi(x) \sim e^{-s \sqrt{x}}$. The qualifier \textit{average} is extremely important here, as the critical wavefunctions would typically exhibit multiple minima and maxima due to the fluctuations in the winding number. As we are primarily interested in disorder-averaged low-frequency transport properties, the leading stretched-exponential part of the wavefunction profile is sufficient.

\section{Mott resonances in the bond-diluted SSH chain}
\label{appb}

In this section, we derive the hybridization energy of the zero modes analytically in the bond-diluted SSH chain model. We, therefore, consider \eqref{sshhda}, with the hoppings missing with a probability $p$ from the unperturbed chain with parameters $t$, $t'$.

\begin{figure}[t!]
        \centering
        \includegraphics[width=0.4\textwidth]{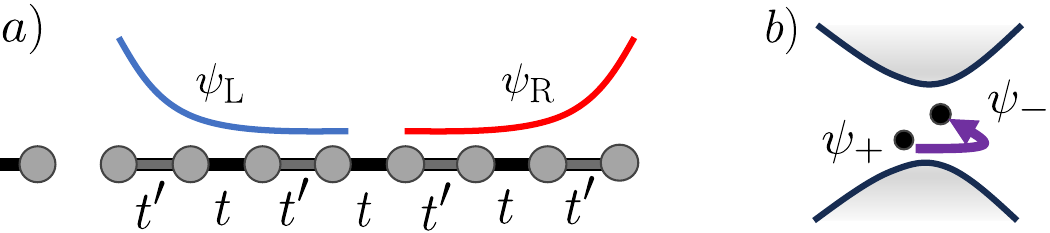}
    \caption{$a)$ Two zero modes ($\psi_{\rm L}$ and $\psi_{\rm R}$) appear on both sides of the topological region created by the bond-dilution disorder in the SSH chain. $b)$ If the segment is sufficiently long, the zero modes hybridize into sub-gap bonding, and anti-bonding states $\psi_{\pm}$. Due to the finite gap, only the optical transitions between such states (indicated with a purple arrow) will contribute to the low-frequency ac conductivity.}
    \label{appplot1}
\end{figure}

As argued in the main text, at low frequency, only the transitions between the hybridized topological zero modes shown in \autoref{appplot1} are relevant for transport. We therefore consider a topological segment of the length $d$, as shown in the image, assuming $|t'|<|t|$. The normalized unperturbed zero modes localized on the ends of the chain take the form
\begin{equation}
\begin{aligned}
\ket{\psi_{\rm L}} & = \sqrt{\frac{2}{\xi_0}} \sum_{0 \leq i \leq d} e^{-i/\xi_0} \ket{i, \rm A} \, , \\
\ket{\psi_{\rm R}} & = \sqrt{\frac{2}{\xi_0}}  \sum_{0 \leq i \leq d} e^{(i-d)/\xi_0} \ket{i, \rm B} \, ,
\end{aligned}
\label{psiab}
\end{equation}
where $\xi_0 = \log^{-1}|t/t'|$, and $d, \xi_0$ assumed large. By repeatedly using the decimation procedure outlined in the next Appendix, the Hamiltonian of the segment is reduced to
\begin{equation}
\hat H_{0} = t' (t'/t)^d \hat c^\dagger_{0\rm A} \hat c_{d\rm B} + {\rm h.c.} \, .
\end{equation}
Note that the parameter $(t'/t)^d$ is always small for sufficiently long sections, which justifies the application of the perturbation theory to obtain the low-lying in energy states. Using degenerate perturbation theory, the energies of the hybridized states $\ket{\psi_{\pm}}$ are found as
\begin{equation}
E_\pm \simeq \pm \braket{\psi_{\rm L}| \hat H_{0} | \psi_{\rm R}} = \pm 2t'/\xi_0 (t'/t)^d =\pm 2t'/\xi_0 e^{-d/\xi_0} \, .
\end{equation}
Therefore, in the bond-diluted SSH chain, the energy of the low-lying Mott resonances depends exponentially on their size $d$.

\section{Real-space renormalization group}
\label{appc}

In this section, we review the real-space renormalization group (RSRG) treatment of the SSH chain with chiral disorder \eqref{sshhda}. The RSRG procedure consists of identifying the largest of the hoppings within the chain (suppose $t_2$) and decimating it using the second-order perturbation theory \cite{dobro}. Consider, e.g., the $t_1 - t_2 - t_3$ cluster of hoppings described by the Hamiltonian
\beq
\hat H = \hat H_{0} + \hat H_{\rm pert}\, , \qquad \hat H_{0} = t_2 \hat c^{\dagger}_2 \hat c_3 + {\rm h.c.} \, , \qquad \hat H_{\rm pert,} = t_1 \hat c^{\dagger}_1 \hat c_2+ t_3 \hat c^{\dagger}_3 \hat c_4 + {\rm h.c.} \, .
\eeq
The unperturbed eigenvalues are $E_0= \pm t_2$ and the eigenstates read
\beq
\ket{\psi_{\pm}} = \frac{1}{\sqrt{2}}(\hat c_2^\dagger \pm \hat c_3^\dagger)\ket{0} \equiv \frac{1}{\sqrt{2}}(\ket{2} \pm \ket{3}) \, .
\eeq
Within the second-order perturbation theory, the effective hopping that replaces the given string $t_1 - t_2 - t_3$ is obtained as
\beq
t_{\rm eff} = \frac{\braket{1|\hat H_{\rm pert}|\psi_{+}}\braket{\psi_{+}|\hat H_{\rm pert}|4}}{t_2} - \frac{\braket{1|\hat H_{\rm pert}|\psi_{-}}\braket{\psi_{-}|\hat H_{\rm pert}|4}}{t_2} = \frac{t_1 t_3}{t_2} \, ,
\label{deci}
\eeq
since
\beq
\bal
\braket{1|\hat H_{\rm pert}|\psi_{+}} & = \frac{t_1}{\sqrt{2}}\, , \qquad
\braket{1|\hat H_{\rm pert}|\psi_{-}} = \frac{t_1}{\sqrt{2}} \, , \\
\braket{4|\hat H_{\rm pert}|\psi_{+}} & = \frac{t_3}{\sqrt{2}} \, , \qquad
\braket{4|\hat H_{\rm pert}|\psi_{-}} = - \frac{t_3}{\sqrt{2}} \, .
\eal
\eeq

The procedure \eqref{deci} is repeated multiple times until the maximum hopping in the chain falls below the cutoff $\Omega$. If we collectively label all odd links $t'$ and all even links $t$, their distributions $P_{t} \equiv P_{t}(t, \Omega)$ and $P_{t'} \equiv P_{t'}(t',\Omega)$ will ``run'' with the cutoff according to
\begin{equation}
\begin{aligned}
-\frac{\partial P_{t}}{\partial \Omega} & =\left[P_t^\Omega-P_{t'}^\Omega \right] P_t+P_{t'}^\Omega \int d t_1 d t_2 \, P_t (t_1, \Omega) P_t (t_2, \Omega) \delta \left(t-t_1 t_2 / \Omega \right), \\
-\frac{\partial P_{t'}}{\partial \Omega} & =\left[P_{t'}^\Omega-P_{t}^\Omega \right] P_{t'}+P_t^\Omega \int d t_1' d t_2' \, P_{t'} (t_1', \Omega) P_{t'} (t_2', \Omega) \delta\left(t'-t_1' t_2' / \Omega\right) \, ,
\end{aligned}
\end{equation}
where $P_t^\Omega \equiv P_t(\Omega, \Omega)$, and $P_{t'}^\Omega \equiv P_{t'}(\Omega, \Omega)$. This system of integro-differential equations was solved in both critical and off-critical cases \cite{fisher1, fisher2}. We review these solutions and their implications below.\\\\
\textit{\underline{Off-critical case:}}\\\\
For convenience, we introduce logarithmic variables 
\begin{equation}
\Gamma \equiv \log(\Omega_0/\Omega) \, , \qquad T \equiv \log(\Omega/t) \, , \qquad T' \equiv \log(\Omega/t') \, .
\end{equation}
If the hoppings $t$ are on average larger than $t'$, the fixed point is characterized by distributions 
\begin{equation}
P_{t}(T) =2 |\delta| \exp [-2 |\delta| T] \, ,  \qquad 
P_{t'}(T') =2 |\delta| e^{-2 |\delta| \Gamma} \exp \left[-2 |\delta| e^{-2 |\delta| \Gamma} T' \right] \, ,
\label{offcdists}
\end{equation}
i.e., the distribution for $T$ (and therefore $t$) becomes cutoff-independent, while the distribution of $T'$ becomes extremely broad already on a logarithmic scale. It implies that extremely small and isolated in energy hoppings $t'$ appear, giving rise to low-energy Mott-like resonances formed by the zero modes hybridizing through small $t'$ bonds. The quantity $\delta$ in the expressions  above is introduced as
\begin{equation}
\delta \equiv \frac{\llangle T \rrangle - \llangle T' \rrangle}{{\rm var}\,T + {\rm var} \, T'} \, .
\end{equation}

The differential equation describes the number of surviving unit cells at the scale $\Gamma$
\begin{equation}
d N = - N (P_t + P_t') d \Gamma \, ,
\label{dNeq}
\end{equation}
as the act of decimation always removes one unit cell. Integration of this equation yields $N \sim \Omega^{- 2 |\delta|}$. The low-energy density of states at energy $E$ is found as $\nu (E) \sim d N^{-1}(E)/ d E \sim E^{2 |\delta| - 1}$. Since almost all decimated hoppings are $t$-hoppings, on the approach to the fixed point, the $t'$ hoppings become typically very long. Their lengths are therefore approximately equal to the typical length of the unit cell, i.e., $d_{t'} \sim N^{-1} \sim \Omega^{2 |\delta|}$. The typical $t'$ hopping itself can be found using the second distribution in \eqref{offcdists}, as 
\begin{equation}
t_{\rm typ}' \sim \exp (\braket{T'}_{\rm D}) \sim \exp(-\Omega^{2 |\delta|}) \sim e^{- d_{t'}} \, .
\end{equation}
Therefore, the hybridization energy of the zero modes across such a hopping element is exponentially small in its length. \\\\
\textit{\underline{Critical case:}}\\\\
The distributions of $T$ and $T'$ \begin{equation}
P_t(T)=\frac{1}{\Gamma} e^{-T / \Gamma}\,  , \qquad P_{t'}(T')=\frac{1}{\Gamma} e^{-T' / \Gamma}
\label{rgcrit}
\end{equation}
both become extremely broad at a low cutoff. The low-energy Mott resonances in this case arise on both $t$ and $t'$ hoppings. Solving \eqref{dNeq} in this case, yields the scaling of the remaining number of unit cells $N \sim \log^{-2} \Omega$, and therefore, $d_{t'} \sim d_t \sim \log^2 \Omega$ due to the $t \leftrightarrow t'$ duality manifest in \eqref{rgcrit}. The low-energy density of states is described by the Dyson singularity $\nu(E) \sim d N(E) / dE \sim (E \log^3 E)^{-1}$. Lastly, using \eqref{rgcrit}, we establish that the remaining at low $\Omega$ hoppings are typically stretched-exponentially dependent on their lengths
\begin{equation}
t_{\rm typ}' \sim t_{\rm typ} \sim \exp (\braket{T}_{\rm D}) \sim e^{-\Gamma} \sim e^{- \sqrt{d}} \, .
\end{equation}

We note here that it is useful to define the dynamical critical exponent $z$ as the polynomial part of the scaling of the typical length of the excitation with the time over which it occurs: $d^z \sim (1/\omega)$. From the discussion above, we obtain $z=1/(2|\delta|)$. The polynomial part of the scaling of the density of states at small energies then reads $\nu(E) \sim E^{-1+1/z}$. Since the low-frequency ac conductivity is given by the pair-wise summation over the symmetric in energy transitions, as shown in the main text, the ac conductivity is itself proportional to the density of states, and hence, $\sigma(\omega) \sim \omega^{1/z}$, up to logarithmic corrections, in consistency with $\sigma_{\rm crit} \sim \log (\Omega/\omega)$, when $z = \infty$.

\section{Mapping to spin and effective continuous models}

\label{appd}

The mapping between the $1D$ Ising models and the $1D$ tight-binding models of Fermions is well-known \cite{mckenzie_exact_1996}. The details of this mapping that apply to the case of an SSH chain with chiral disorder are reviewed below. Additionally, at the critical point, the properties of these lattice models are identical to those of the continuous Dirac-fermion model with random mass studied extensively earlier. This allows making a host of exact statements about the properties of the SSH chain at and around criticality.

The starting point is the $XX$ Ising model in $1D$
\begin{equation}
    \hat H_{\rm Ising} = \sum_i J_i [\hat \sigma_i^x \hat \sigma_{i+1}^x + \hat \sigma_i^y \hat \sigma_{i+1}^y] \, ,
    \label{ising}
\end{equation}
where $J_{i\,\rm odd} = (t' - \epsilon_j)/2$ with random $\epsilon_i$ and $J_{i\,\rm even} = t/2$. Using the Jordan-Wigner transformation
\begin{equation}
    \begin{aligned}
    & \hat{\sigma}_i^x \hat{\sigma}_{i+1}^x=\left(\hat{c}_i^{\dagger} \hat{c}_{i+1}+\hat{c}_i^{\dagger} \hat{c}_{i+1}^{\dagger}+\text{h.c.}\right)\, , \\
    & \hat{\sigma}_i^y \hat{\sigma}_{i+1}^y=\left(\hat{c}_i^{\dagger} \hat{c}_{i+1}-\hat{c}_i^{\dagger} \hat{c}_{i+1}^{\dagger}+\text {h.c.}\right)\, ,
\end{aligned}
\end{equation}
it is straightforward to show that the Hamiltonian \eqref{ising} is equivalent to that for the SSH chain with chiral disorder studied in the main text. The dynamical spin conductivity in the model \eqref{ising} was studied in \cite{motr}, where it was shown that away from the criticality (random dimer phase)
\begin{equation}
    \sigma(\omega) \sim s^{-2} \omega^{1/z} \log^2 \omega \, ,
\end{equation}
where $z$ is the dynamical critical exponent. At criticality, \cite{motr} find
\begin{equation}
    \sigma_{\rm crit}(\omega) \sim s^{-2} \log \omega \, .
\end{equation}
These expressions agree with those in the main text once we set $z = (2 |\delta|)^{-1}$, as shown below. It is noteworthy that the low-frequency spin transport in this case arises from hybridized domain wall states on both sides of the rare regions created by disorder in magnetic couplings. For example, these rare regions may arise from the extensive incorporation of the even-bond-ordered phase into a predominantly odd-bond-ordered environment. Such hybridized domain walls are non-local duals of the Mott resonances described in the main text.

Some further exact analytical results for the SSH chain follow from the low-energy theory for the SSH chain Hamiltonian (see main text) expanded around $k_{\rm F} = \pi$
\begin{equation}
    \hat H_{\rm cont} = \sum_x \hat c^\dagger_x [ i v_{\rm F}  \hat \sigma_2 \prl_x + (t' - t - \varepsilon_x) \hat \sigma_1]\,  \hat c_x \, ,
\end{equation}
where $v_{\rm F} = t$. In the extensive studies of this model summarized in \cite{mckenzie_exact_1996} it was established that away from criticality, $\nu(E) \sim E^{-1+2|\delta|}$ and the dynamical critical exponent $z=(2 |\delta|)^{-1}$. At criticality ($\delta=0$), the density of states takes the form of the Dyson singularity $\nu(E) \sim s^2 [|E| \log^3 (s^2/|E|)]^{-1}$. The results found within this continuum model apply to the full tight-binding case when the dimerization is small, and the disorder is low, i.e., $t \gg |t-t'|$, $s^2 \ll 1$.

\section{Variable-range hopping and temperature-dependent dc conductivity}
\label{appe}

In this appendix, we obtain the temperature dependence of the dc conductivity due to variable-range hopping between the topological zero modes in the SSH chain with bond disorder \eqref{sshhda}.

Consider two off-critical $|\delta| > 0$ zero-mode states $\psi_1$ and $\psi_2$ centered at the positions $x_1$ and $x_2$, and separated in the real space by $d=|x_1-x_2|$. The phonon-assisted hopping rate between these localized states is governed by the competition between the exponentially decaying wavefunction overlap and the Boltzmann weight \cite{shk},
\begin{equation}
W(x_1,x_2)\sim |\langle \psi_1|\hat x|\psi_2 \rangle|^2 \,
e^{-\frac{\Delta E}{k_{\rm B}T}} .
\label{wij}
\end{equation}
The typical energy spacing between states within a segment of length $d$ is determined by the condition that there is $O(1)$ state in that window. Away from criticality in the random-bond SSH chain model, where the density of states scales as $\nu(E)\sim E^{2\delta-1}$, we find
\begin{equation}
d\int_0^{\Delta E} dE\,\nu(E) \sim 1 \Rightarrow \quad
\Delta E\sim d^{-1/(2\delta)} .
\end{equation}
For the regular exponentially localized states with $|\langle \psi_i|\hat x|\psi_j\rangle|^2\sim e^{-2d/\xi}$, the hopping rate becomes
\begin{equation}
W(d)=\exp\!\left(-\frac{2d}{\xi}
-c\frac{d^{-1/(2\delta)}}{k_{\rm B}T}\right)\, ,
\end{equation}
where $c$ is a constant. Minimizing the exponent yields the optimal hopping distance
\begin{equation}
\overline d \sim (T_0/T)^{\frac{2\delta}{1+2\delta}},
\end{equation}
and from \eqref{wij}, we find
\begin{equation}
\sigma(T)\sim W(\overline d)\sim 
\exp\!\left[-(T_0/T)^{\frac{2\delta}{1+2\delta}}\right] \, .
\label{sT}
\end{equation}

For $\delta=1/2$, the density of states is constant and one recovers the standard one-dimensional Mott variable-range hopping law \cite{mott_electronic_2012}, observed, e.g., in polymer nanofibers \cite{vrh}. In contrast, in the limit $\delta\to 0$, the Mott exponent vanishes, indicating the transition between the tunneling transport regime and the polynomial temperature scaling characteristic of a metal. This signals the onset of the activated dynamical scaling, characterized by the divergence of a dynamical critical exponent $z = (2 |\delta|)^{-1}$.

 \bibliography{refs}

% refs.bib is not editable due to size. use refsadd.bib
\end{document}